\documentstyle[prl,twocolumn,aps,psfig]{revtex}
\def\tdot{\stackrel{\dots}}
\def\order{{\cal O}}
\begin{document}
\preprint{TAUP 2549-99}
\twocolumn[\hsize\textwidth\columnwidth\hsize\csname@twocolumnfalse%
\endcsname
\title{The Effect of Radiation on the Stochastic Web}
\author{
Y. Ashkenazy$^{\text{1}}$\footnote{email-address: ashkenaz@mail.biu.ac.il} and 
L. P. Horwitz$^{\text{1,2}}$\footnote{email-address: larry@post.tau.ac.il}}
\address{$^1$ Department of Physics, Bar-Ilan University, Ramat-Gan 52900,
Israel\\
$^2$ School of Physics, Raymond and Beverly Sackler Faculty of Exact Sciences\\
Tel-Aviv University, Ramat-Aviv, Israel.}
\date{\today}
\maketitle
\begin{abstract}
{
A charged particle circling in a uniform magnetic field and kicked by an 
electric field is considered. Under the assumption of small magnetic field, an 
iterative map is developed. Comparison between the (relativistic) non-radiative
case and the (relativistic) radiative case shows that in both cases one can 
observe
a stochastic web structure, and that both cases are qualitatively similar. 
}
\end{abstract}
\pacs{
PACS numbers: XXXXXX}
]
\narrowtext
\newpage
\section{Introduction}
Zaslavskii {\it et al} \cite{Zaslavskii86} studied the behavior of particles 
in the wave packet of an electric field in the presence of a 
static magnetic field.
For a broad wave packet with sufficiently uniform spectrum, one may show that 
the problem can be stated in terms of an electrically kicked harmonic 
oscillator. For rational ratios between the frequency
of the kicking field and Larmor frequency associated with the 
magnetic field the phase space of the system is covered by a mesh of finite 
thickness; inside the filaments of the mesh, the dynamics of the particle is 
stochastic and
outside (in the cells of stability), the dynamics is regular. This structure is
called a stochastic web. It was found that this pattern covers the entire 
phase plane, permitting the particle to diffuse arbitrarily far into the 
region of high energies (a process analogous to Arnol'd 
diffusion \cite{Arnold64}).

Since the stochastic web leads to unbounded energies, several authors have
considered the corresponding relativistic problem. Longcope and Sudan 
\cite{Longcope87} studied this system (in effectively 
$1\frac{1}{2}$ dimensions) and found that for initial conditions close to the
origin of the phase space there is a stochastic web, which is bounded in
energy, of a form quite similar, in the neighborhood of the origin, to the 
non-relativistic case treated by
Zaslavskii {\it et al}. Karimabadi and Angelopoulos \cite{Karimabadi89}
studied the case of an obliquely propagating wave, and showed that under 
certain conditions, particles can be accelerated to unlimited energy through
an Arnol'd diffusion in two dimensions. 
Since an accelerated charged particle radiates, it is 
important to study the radiative corrections to this motion. We shall use the
Lorentz-Dirac equation to compute this effect.

We compute solutions to this equation for the 
case of the kicked oscillator. At low velocities, the stochastic web
found by Zaslavskii {\it et al}\cite{Zaslavskii86} occurs; the system diffuses
in the stochastic region to unbounded energy, as found by Karimabadi and
Angelopoulos\cite{Karimabadi89}. The velocity of the particle is light
speed limited by the dynamical equations, in particular, by the suppression
of the action of the electric field at velocities approaching the velocity 
of light \cite{Goldman}.

\section{Model}
In the present study we will consider a charged particle moving in a uniform 
magnetic field, and kicked by an electric field. The effect of relativity, as 
well
as the radiation of the particle, will be considered. We restrict ourselves to
the ``on mass shell'' case\footnote{The more general case which includes also
the possibility ``off mass shell'' motion, is discussed in a separate article 
in this journal \cite{Horwitz}.}.

The fundamental equation that we use to study radiation is the Lorentz-Dirac
equation \cite{Dirac}, is
\begin{equation} 
\label{e1}
m_0\ddot x^\mu={e \over c}\dot x_\nu F^{\mu\nu} + \gamma_0m_0({\tdot x^\mu}-
{1 \over c^2} \dot x^\mu \ddot x_\nu \ddot x^\nu),
\end{equation}
where $\gamma_0={2 \over 3}{r_0 \over c}=6.26\times 10^{-24}\sec$. $\mu$ and
$\nu$ are the coordinates $t$ and $x$, $y$, and $z$ (or 0, 1, 2, 3),
and the derivative is
with respect to $\tau$, which can be regarded as proper time in the 
``on-mass-shell'' case. $F^{\mu\nu}$ is the antisymmetric electromagnetic
tensor. The first term of the right hand side of Eq. (\ref{e1}) is the 
relativistic Lorentz force, while the second term is the radiation-reaction 
term. Note that the small size of the radiation coefficient 
($\gamma_0 \ll 1$) leads to a singular equation that requires special 
mathematical treatment, as well as physical restrictions, as will be shown in
the succeeding sections.

Following Zaslavskii {\it et al} \cite{Zaslavskii86}, the magnetic field is 
chosen to be a uniform field in the $z$ direction, and the kicking
electric field is a function of $x$ in the $x$ direction,
\begin {eqnarray}
\label{e2}
\vec B &=& (0,0,B_0) \nonumber \\
\vec E(x,t) &=& ({f(x){\sum_{n=-\infty}^\infty\delta(t-nT)}},0,0).
\end{eqnarray}
Originally, Zaslavskii {\it et al} chose a uniform 
broad band electric field wave packet which can be expanded as an infinite sum 
of (kicking) $\delta$-functions ($\omega_0$ is the frequency of the central 
harmonic of the wave packet, $k_0$ is the wave number of the central harmonic, 
and $\Delta \omega$ is the frequency between the harmonics of the wave packet)
\begin{equation}
\label{e3a}
E_x=E(x,t)=-E_0\sum_{n=-\infty}^\infty \sin (k_0x-\omega_0t-n\Delta\omega t),
\end{equation}
which, for $\omega_0=0$, becomes
\begin{equation}
\label{e3}
E(x,t)= -E_0T\sin(k_0x)\sum_{n=-\infty}^\infty\delta(t-nT),
\end{equation}
where $T={{2\pi}\over{\Delta\omega}}$.
Eq. (\ref{e2}) is the generalization of Eq. (\ref{e3}), with an
arbitrary function $f(x)$ instead of the sine function of Eq. (\ref{e3}).

In order to introduce a map which connects between cycles of integration
which start before a kick and end before the next electric field
kick, one has first to integrate over the $\delta$ electric kick and then 
to integrate the equations of motion between the kicks, where only a uniform 
magnetic field is present and there is no electric field up to the beginning 
of the next kick.

\section{The motion of a charged particle in a uniform magnetic field}
In the case of a uniform magnetic field from Eq. (\ref{e2}), Eqs. (\ref{e1})
reduce to three coupled differential equations,
\begin{eqnarray}
\label{e4}
c \ddot t &=\ddot x_0=& \gamma_0({\tdot x_0} - {1 \over {c^2}} 
\dot x_0R) \nonumber \\
\ddot x &=\ddot x_1=& -\Omega \dot x_2 + \gamma_0({\tdot x_1} - 
{1 \over {c^2}} \dot x_1R) \\
\ddot y &=\ddot x_2=& \Omega \dot x_1 + \gamma_0({\tdot x_2} - 
{1 \over {c^2}} \dot x_2R), \nonumber
\end{eqnarray}
where $\Omega={{e_0B_0} \over {m_0c}}$, 
$R=\ddot x_1^2+\ddot x_2^2-\ddot x_0^2$, and $e_0=-e$ (charge of the electron).

Using a complex coordinate\cite{Sokolov}, $u=x+iy$, Eqs. (\ref{e4}) can be 
written as,
\begin{eqnarray}
\label{e5}
\ddot t &=& \gamma_0({\tdot t}-{1 \over {c^2}} \dot t R) \nonumber \\
\ddot u &=& i\Omega\dot u+\gamma_0({\tdot u}-{1 \over {c^2}} \dot u R).
\end{eqnarray}
We wish here to constrain the radiatively perturbed motion to the ``mass 
shell''; to do this, it is very convenient to use hyperbolic coordinates,
\begin{eqnarray}
\label{e6}
\dot t &=& \cosh q \nonumber \\
\dot x &=& c \sinh q \cos \phi \\
\dot y &=& c \sinh q \sin \phi \nonumber
\end{eqnarray}
since the ``on mass shell'' restriction (see \cite{Horwitz} for further 
discussion),
\begin{equation}
\label{e7}
\dot x^\mu \dot x_\mu = \dot x^2+\dot y^2-c^2 \dot t^2 = -c^2,
\end{equation}
is then automatically satisfied. The complex coordinate $u$ then becomes,
\begin{equation}
\label{7a}
\dot u = c \sinh q e^{i\phi}
\end{equation}
and the equations of motion Eqs. (\ref{e5}) can be written as,
\begin{eqnarray}
\dot q &=& -\gamma_0 \dot\phi^2 \cosh q\sinh q + \gamma_0\ddot q \label{e8}\\
\dot \phi &=& \Omega +2\gamma_0\dot\phi\dot q \coth q+\gamma_0 \ddot \phi 
\label{e9}.
\end{eqnarray}
As pointed out earlier, in the case of a singular equation (such as Eqs. 
(\ref{e8}) and (\ref{e9})), one has to consider physical arguments as well as mathematical arguments. There are several suggested methods to avoid the 
``run away electron'' problem \cite{Dirac}. One of the most frequently used, 
especially useful for scattering problems, assumes \cite{Rohrlich} that the 
particle loses all its energy after a sufficiently large time, and the run 
away parts of the solution are set to zero; the equations
 of motion can then be written as an integral equation. Another method, 
which permits the study of problems with only non-asymptotic states 
\cite{Aguirrebiria},
uses an iterative singular perturbation integration that leads to the stable 
solution. Since the  integration we must use is over a finite time,
 it is 
impossible to use an asymptotic condition (e.g. $|v| \to 0$ at $\tau \to 
\infty$).
The Sokolov and Ternov approach \cite{Sokolov} is more suitable for our
purpose, since it can be implemented for bounded time integration, and the 
mathematical solution is simple. In this method, in the first step,
the perturbation terms in Eq. (\ref{e9}) are disregarded (as in the iterative 
scheme of \cite{Aguirrebiria}) , and then the resulting  
equation is exactly integrated. In the second step, the singular term on 
the right hand side of 
Eq. (\ref{e8}) is not considered; using $\phi (\tau)$ from Eq. (\ref{e9}), the
equation including just the first term is integrated. The solution is 
\cite{Sokolov}
\begin{eqnarray}
\label{e10}
\phi &=& \Omega \tau \nonumber \\
\beta(\tau) &=& \tanh q = {v_0 \over c} e^{-\gamma_0\Omega^2 \tau} = 
\beta_0 e^{-{\tau \over \tau_0}},
\end{eqnarray}
where $\beta_0={v_0 \over c}$ is the actual (normalized) velocity, and 
$\tau_0={1 \over {\gamma_0\Omega^2}}$ is the decay time for the energy of the particle.

To get an estimate for the radiation during one cycle, we will 
refer to the maximal uniform magnetic field that can achieved today in a 
laboratory, which is\footnote{It 
is possible to obtain a short-lived magnetic field of 
$\order(10^6\,{\rm Gauss})$ in the laboratory.} 
$\order(10T) = \order(10^5\, {\rm Gauss})$. If, for example, 
$B_0=10^5\, {\rm Gauss}$, then $\Omega=1.76\times10^{12}{1 \over \sec}$, and 
$\tau_0=5\times 10^{-2} \sec$. Thus, it is clear from Eq. (\ref{e10}) that the
particle makes ${\Omega \over {2\pi}}\tau_0 \approx 10^{10}$ cycles before
it decays to ${1 \over e}$ of its initial velocity. In other words, the energy
loss during one cycle in very small, and since in our problem, the time $T$
between the kicking is of the order of the period, the energy loss 
between consecutive kicks is very small.  At much higher field strengths, a 
qualitatively different behavior may occur, for which the energy loss may be 
very high during a single cycle, and the resulting motion may be modified to 
reflect a nonrelativistic behavior, or it may entirely stop before the next 
kick. This behavior 
may result in a stochastic web of a somewhat different type.

\begin{figure}[thb]
\psfig{figure=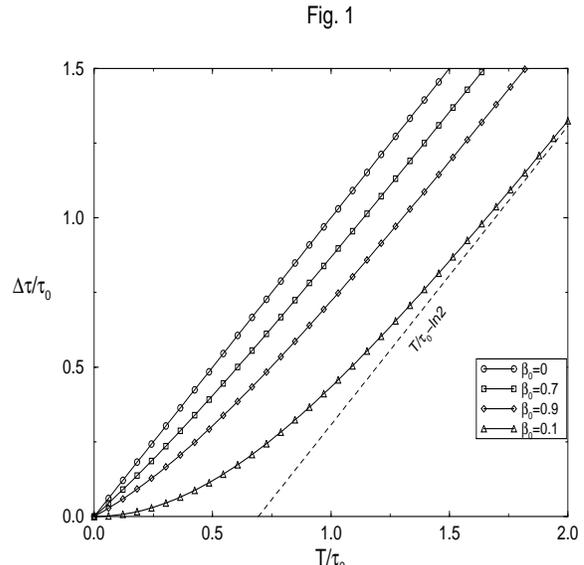,height=8cm,width=8.5cm,angle=-90}
\caption{\label{fig1}
The function ${{\Delta \tau (T / \tau_0)}\over{\tau_0}}$ versus 
$T / \tau_0$. Four typical 
cases are shown ($\beta_0=0$, $\beta_0=0.7$, $\beta_0=0.9$, $\beta_0=1$).
All possible curve lines between the graphs of $\beta_0=0$ and $\beta_0=1$.
The dashed line indicates the asymptotic behavior of the case $\beta_0=1$.
}
\end{figure}

The time between the kicking is measured according to the observed time along the motion, 
$\Delta t = T$. Thus, it is necessary to find the corresponding $\Delta \tau$.
It follows from Eq. (\ref{e6}) and Eq. (\ref{e10}) that,
\begin{equation}
\label{e11}
\dot t = \cosh q = {1 \over \sqrt{1-\beta^2_0 e^{-{{2\tau} \over {\tau_0}}}}}.
\end{equation}
The solution of Eq. (\ref{e11}) can be obtained by a elementary integration,
\begin{equation}
\label{e12}
t=\tau_0\ln \left({{1+\sqrt{1-\alpha^2}} \over \alpha}C_1\right),
\end{equation}
where $\alpha=\beta_0 e^{-{\tau \over {\tau_0}}}$. After some algebraic 
operations one gets,
\begin{equation}
\label{e13}
\Delta \tau = \tau_0 \ln \left( {1 \over 2}{{\beta_0^2+(1+\sqrt{1-\beta_0^2})^2
e^{2T/\tau_0}} \over {(1+\sqrt{1-\beta_0^2})e^{T/\tau_0}}} \right).
\end{equation}
In Fig. \ref{fig1} we present the behavior of the function 
${{\Delta \tau (T / \tau_0)}\over{\tau_0}}$. It can be seen from Fig. 
\ref{fig1} that in any case,
\begin{equation}
\label{e14}
\Delta \tau \le T;
\end{equation}
this implies  that the time difference according to the
 proper time $\tau$ is always less then the time difference according 
to observed time $t$. This fact is similar to the well known relativistic 
``time-dilation'' phenomenon. The inequality Eq. (\ref{e14})
can be shown also by considering the two extreme cases,
\begin{eqnarray}
T \gg \tau_0 & \Rightarrow & \Delta \tau \approx T+\tau_0 \ln 
\left({{1+\sqrt{1-\beta_0^2}} \over 2} \right) \label{e15} \\
T \ll \tau_0 & \Rightarrow & \Delta \tau \approx T\sqrt{1-\beta_0^2}
+{{\beta_0^2} \over 2}{{T^2} \over {\tau_0}}+ \order(T^3) \label{e16}.
\end{eqnarray}
For strong magnetic field (i.e. the case of Eq. (\ref{e15})) the numerator is
less (or equal) then 2 and thus the $ln$ function will give a negative number;
in this case the inequality Eq. (\ref{e14}) is clearly achieved.
For a low magnetic field Eq. (\ref{e16}) behaves like a parabola; 
the slope for $T=0$
is $\sqrt{1-\beta_0^2}$ which is less than one, and thus the inequality Eq. 
(\ref{e14}) is satisfied.
As explained above, in the present study we confine ourselves to the case of
very small $T$, since this time difference is constrained because of the
limited magnetic field that can achieved in the laboratory. Thus, just Eq.
(\ref{e16}) is applicable in this study.

\section{Derivation of the map}
In the previous section we have shown the approximate solution for a charged
particle circling in a uniform magnetic field. However, as mentioned above,
 the kicking of
the electric field from Eq. (\ref{e2}) is with respect to the observed time 
$t$, and it is therefore convenient to integrate the equations of motion 
(Eqs. (\ref{e8}) and (\ref{e9})) with respect to $t$.

\subsection{A charged particle in a uniform magnetic field - integration with 
respect to the observed time $t$}
The arguments that were used in order to obtain Eqs. (\ref{e10}),
as well as the implementation of the chain rule on Eqs. (\ref{e8}) and 
(\ref{e9}) by the use of Eq. (\ref{e6}) lead to the following equations of 
motion
\begin{eqnarray}
{{dq}\over{dt}} &=& -{1\over{\tau_0}}\sinh q \label{e17} \\
{{d\phi}\over{dt}} &=& {\Omega \over {\cosh q}} \label{e18}.
\end{eqnarray}
A simple integration of Eq. (\ref{e17}) yields,
\begin{equation}
\label{e19}
\tanh {q\over 2} = e^{C_1}e^{-{t \over{\tau_0}}},
\end{equation}
where $C_1<0$.
Substitution of Eq. (\ref{e19}) in Eq. (\ref{e18}) gives,
\begin{equation}
\label{e20}
{{d\phi}\over{dt}} = \Omega \tanh ({t\over{\tau_0}}-C_1),
\end{equation}
and the solution is
\begin{equation}
\label{e21}
\phi = \Omega\tau_0\ln\cosh({t\over {\tau}}-C_1)+C_2,
\end{equation}
where $C_1=\cosh^{-1}({1\over{\beta_0}})$, and $C_2=\tan^{-1} 
(\dot y_0 / \dot x_0)+\Omega\tau_0\ln\beta_0$. 
Returning to the original $x$, $y$ coordinates (using Eqs. (\ref{e6})),
the equations of motion become,
\begin{eqnarray}
{{dx} \over {dt}} &=& c {{\cos(\Omega\tau_0\ln(\cosh(t/\tau_0-C_1))+C_2)} \over
{\cosh(t/\tau_0-C_1)}} \nonumber \\
{{dy} \over {dt}} &=& c {{\sin(\Omega\tau_0\ln(\cosh(t/\tau_0-C_1))+C_2)} \over
{\cosh(t/\tau_0-C_1)}} \label{e22}.
\end{eqnarray}
It is possible to write Eq. (\ref{e22}) in more convenient way, by using
the relation
\begin{equation}
\label{e23}
\cosh({t \over {\tau_0}}-C_1)={1 \over {\beta_0}} \cosh{t \over {\tau_0}}
+{1 \over {\beta_0}}\sqrt{1-\beta_0^2}\sinh{t \over {\tau_0}}.
\end{equation}
Eqs. (\ref{e22}) then become
\begin{eqnarray}
{{dx} \over {dt}} &=& {{\left({{dx}\over{dt}}\right)_0\cos\alpha-
\left({{dy}\over{dt}}\right)_0\sin\alpha} \over {\cosh{t\over{\tau_0}}+
\sqrt{1-\beta_0^2}\sinh{t\over{\tau_0}}}} \nonumber \\
{{dy} \over {dt}} &=& {{\left({{dx}\over{dt}}\right)_0\sin\alpha+
\left({{dy}\over{dt}}\right)_0\cos\alpha} \over {\cosh{t\over{\tau_0}}+
\sqrt{1-\beta_0^2}\sinh{t\over{\tau_0}}}} \label{e24},
\end{eqnarray}
where
\begin{equation}
\label{e25}
\alpha = \Omega\tau_0\ln (\cosh{t\over{\tau_0}}(1+\sqrt{1-\beta_0^2}
\tanh{t\over{\tau_0}})).
\end{equation}

It is necessary to integrate Eq. (\ref{e24}) since the value of $x$ is used in 
the electric field kicking in Eq. (\ref{e2}). There is no analytical 
solution to Eq. (\ref{e24}), and a numerical integration must, in general,
be performed. 
However, since we restrict ourselves to $T/\tau_0\ll 1$, it is possible to 
expand ${{dx}\over{dt}}$ and ${{dy}\over{dt}}$ in a Taylor series and then 
to integrate. The expansion of Eq. (\ref{e25}) is
\begin{equation}
\label{e26}
\alpha \approx \Omega \sqrt{1-\beta_0^2}t
+{1 \over 2}\beta_0^2{{t^2} \over {\tau_0}}+{1\over 3}\sqrt{1-\beta_0^2}
{{t^3} \over {\tau_0^2}}({3\over 2}-\beta_0^2)+ 
\order({{{t^4} \over {\tau_0^3}}}).
\end{equation}
The first order expansion (according to $t/\tau_0$) of Eqs. (\ref{e24}) is
\begin{eqnarray}
{{dx} \over {dt}} = {1\over{1+{1\over\gamma}{t\over{\tau_0}}}} &&
\left[\left({{dx}\over{dt}}\right)_0\cos({{\Omega}\over\gamma}t+{1\over2}
\Omega\beta_0^2{{t^2}\over{\tau_0}}) \right. \nonumber \\
&& \left. -\left({{dy}\over{dt}}\right)_0\sin({{\Omega}\over\gamma}t+{1\over2}
\Omega\beta_0^2{{t^2}\over{\tau_0}})\right] \nonumber \\
{{dy} \over {dt}} = {1\over{1+{1\over\gamma}{t\over{\tau_0}}}} &&
\left[\left({{dx}\over{dt}}\right)_0\sin({{\Omega}\over\gamma}t+{1\over2}
\Omega\beta_0^2{{t^2}\over{\tau_0}})  \right. \nonumber \\
&& \left. +\left({{dy}\over{dt}}\right)_0\cos({{\Omega}\over\gamma}t+{1\over2}
\Omega\beta_0^2{{t^2}\over{\tau_0}})\right], \label{e27}
\end{eqnarray}
where $\gamma=1/\sqrt{1-\beta_0^2}$.
Since ${1\over\gamma}{t\over{\tau_0}}\ll 1$, one can use the approximation,
\begin{equation}
\label{e28}
{1\over{1+{1\over\gamma}{t\over{\tau_0}}}} \approx 
{{1-{1\over\gamma}{t\over{\tau_0}}}}.
\end{equation}
Using Eq. (\ref{e28}), the actual velocities, $dx/dt$ and $dy/dt$, from Eq. 
(\ref{e27}) can be integrated and expressed by elementary Fresnel functions.
However, the effect of radiation is due to the ${1\over\gamma}{t\over{\tau_0}}$
term (which multiplies the sine and cosine functions), and the 
${1\over2}\Omega\beta_0^2{{t^2}\over{\tau_0}}$ term is not essential since the
major offset from the unstable fixed point is due to the 
${{\Omega}\over\gamma}t$ term. 

Under the above assumptions the resulting equations are,
\begin{eqnarray} 
\left({{dx} \over {dt}}\right)_T &=& \left(1-{T\over{\gamma\tau_0}}\right) 
\times \nonumber \\
&& \left[\left({{dx}\over{dt}}\right)_0\cos({{\Omega}\over\gamma}T)-
\left({{dy}\over{dt}}\right)_0\sin({{\Omega}\over\gamma}T)\right] \label{e29}\\
\left({{dy} \over {dt}}\right)_T &=& \left(1-{T\over{\gamma\tau_0}}\right)
\times \nonumber \\
&& \left[\left({{dx}\over{dt}}\right)_0\sin({{\Omega}\over\gamma}T)+
\left({{dy}\over{dt}}\right)_0\cos({{\Omega}\over\gamma}T)\right]. \label{e30}
\end{eqnarray}
One therefore obtains
\begin{equation}
x_T \approx {\gamma \over \Omega}\left[\left({{dy}\over{dt}}\right)_T
-\left({{dy}\over{dt}}\right)_0\right]-
{\gamma \over {\Omega^2\tau_0}}\left[\left({{dx}\over{dt}}\right)_T
-\left({{dx}\over{dt}}\right)_0\right]+x_0. \label{e31}
\end{equation}
The exponential decay from Eqs. (\ref{e22}) are replaced by linear decay.

\subsection{Integration over a $\delta$ electric kick}
As pointed out in the previous section, the Lorentz-Dirac equation 
(Eq. (\ref{e1}))
is a singular equation. The electric field which is used in this paper was 
expanded to a sum of $\delta$ functions (Eq. (\ref{e2})). In that case, Eq. 
(\ref{e1}) can not be integrated over the kick by regular treatment, and some 
approximation
for the $\delta$ function should be considered instead. In the present paper 
we assumed that the radiation during the kick is negligible; the radiation
during the kick will be studied elsewhere. 
In addition, since the integration is over an infinitesimal time interval, 
there is no need to consider the constant magnetic field during the kick.

Under the above assumptions, in the neighborhood of the kick, 
the Lorentz-Dirac equation, Eq. (\ref{e1}), can 
be written as,
\begin{eqnarray}
{d\over{dt}}\left({{dx}\over{d\tau}}\right) &=& 
{f(x){\sum_{n=-\infty}^\infty\delta(t-nT)}} \nonumber \\
{d\over{dt}}\left({{dy}\over{d\tau}}\right) &=& 0.\label{e32}
\end{eqnarray}
Integration over the $\delta$ function yields,
\begin{eqnarray}
\left({{dx}\over{d\tau}}\right)_+ &=& \left({{dx}\over{d\tau}}\right)_- 
+ f(x) \nonumber \\
\left({{dy}\over{d\tau}}\right)_+ &=& \left({{dy}\over{d\tau}}\right)_-
\label{e33}
\end{eqnarray}
were the + sign indicates after the kick, and the - sign, before the kick.
Following Ref. \cite{Zaslavskii86}, we chose,
\begin{equation}
\label{e34}
f(x) = {{eET}\over{m_0}}\sin(kx) = K\sin(kx).
\end{equation}
Using the fact that ${d\over{d\tau}}=\gamma{d\over{dt}}$ one obtains the 
following relations,
\begin{eqnarray}
\gamma_+ &=& \sqrt{1+{1\over{c^2}}\left[\left(\gamma_-\left({{dx}\over{dt}}
\right)_-+f(x_n)\right)^2+\gamma_-^2\left({{dy}\over{dt}}\right)_-^2\right]}
\nonumber \\
&&\left({{dx}\over{dt}}\right)_+ = {\gamma_-\over\gamma_+}\left({{dx}\over{dt}}
\right)_-+{1\over\gamma_+}f(x_n) \label{e35} \\
&&\left({{dy}\over{dt}}\right)_+ = {\gamma_-\over\gamma_+}\left({{dy}\over{dt}}
\right)_-\nonumber.
\end{eqnarray}
Returning to the initial charge $e=-e_0$ and thus $\Omega \to -\Omega$, the map
which connects the velocities from just before the kick to the next kick
(by the use of Eqs. (\ref{e29})-(\ref{e31})) is,
\begin{eqnarray} 
\left({{dx} \over {dt}}\right)_{n+1}&&=\left(1-{T\over{\gamma_+\tau_0}}\right)
\times \nonumber \\
&& \left[\left({{dx}\over{dt}}\right)_+\cos({{\Omega}\over{\gamma_+}}T)
+\left({{dy}\over{dt}}\right)_+\sin({{\Omega}\over{\gamma_+}}T)
\right] \label{e36}\\
\left({{dy} \over {dt}}\right)_{n+1}&&=\left(1-{T\over{\gamma_+\tau_0}}\right)
\times \nonumber \\
&& \left[-\left({{dx}\over{dt}}\right)_+\sin({{\Omega}\over{\gamma_+}}T)
+ \left({{dy}\over{dt}}\right)_+\cos({{\Omega}\over{\gamma_+}}T)
\right] \label{e37}\\
x_{n+1} &&\approx -{{\gamma_+} \over \Omega}\left[\left({{dy}\over{dt}}\right)_
{n+1}-\left({{dy}\over{dt}}\right)_+\right] \nonumber \\
&&-{{\gamma_+} \over {\Omega^2\tau_0}}\left[\left({{dx}\over{dt}}\right)_{n+1}
-\left({{dx}\over{dt}}\right)_+\right]+x_n. \label{e38}
\end{eqnarray}
Notice that the minus signs in Eq. (\ref{e35}) refer to the $n^{th}$ points of
the map.

It is possible to return to the non-radiative limit by letting 
${1\over{\tau_0}}\to 0$. In that case the map is equivalent to the map which 
was derived by Longcope and Sudan \cite{Longcope87}. The nonrelativistic limit,
which was derived by Zaslavskii {\it et al} \cite {Zaslavskii86}, 
is achieved by letting ${1\over{\tau_0}}\to 0$ and $c\to \infty$.

\subsection{Analysis}
\begin{figure}
\psfig{figure=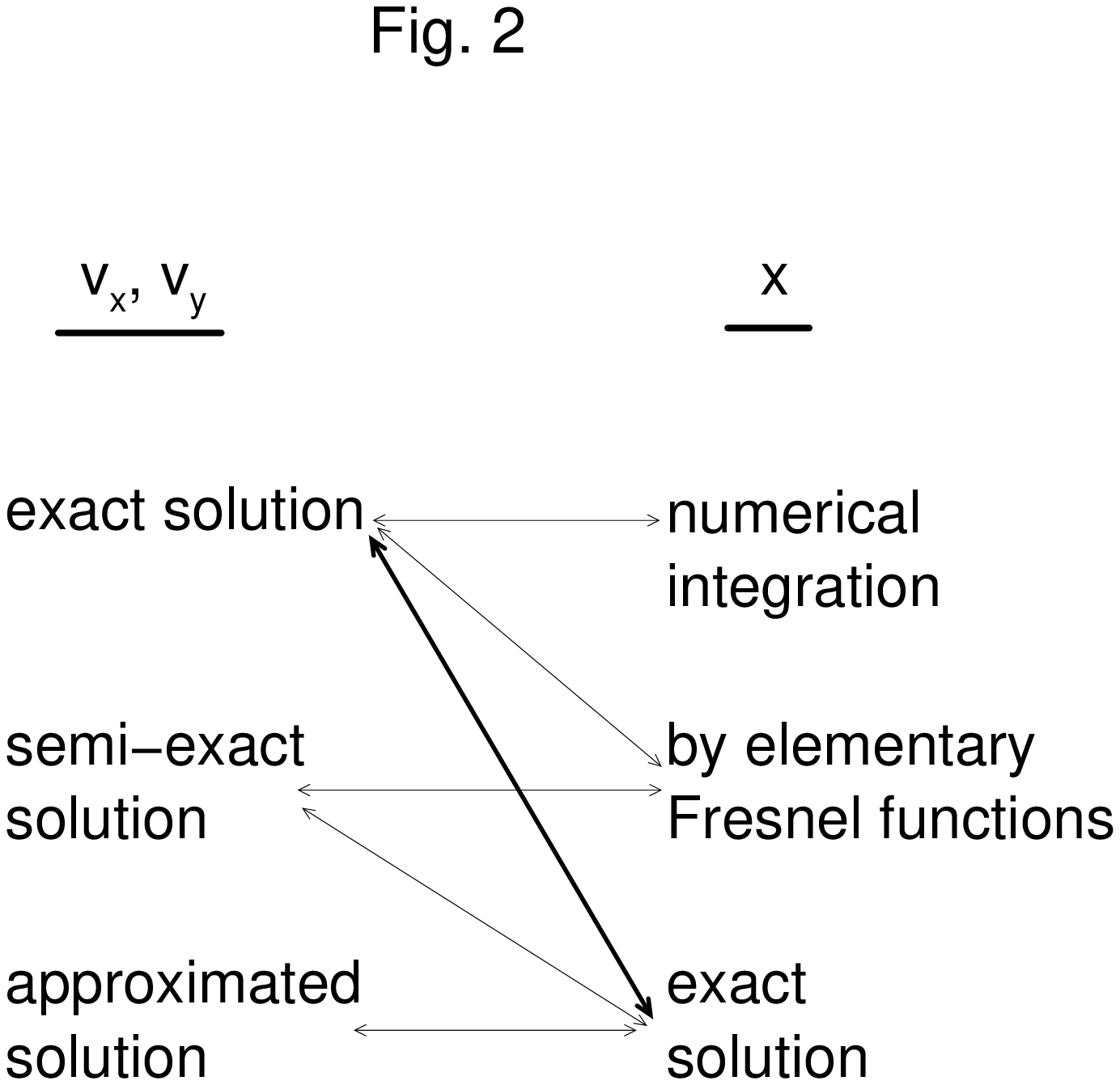,height=10cm,width=10.5cm,angle=-90}
\caption[tbp]{\label{fig2}
A diagram which present the all possibilties for the map construction. 
The wider line indicates the chosen combination.
}
\end{figure}

Up to the derivation of Eqs. (\ref{e36})-(\ref{e38}) we have passed three 
stages, (a) the velocities were calculated using an exact solution and the 
position $x$ can be calculated by numerical integration (Eqs. (\ref{e24})), (b)
the velocities were approximated by Eqs. (\ref{e27}) and the position $x$ by
the elementary Fresnel function, and (c) the velocities were approximated by 
Eqs. (\ref{e36})-(\ref{e37}) and the position $x$ (Eq. (\ref{e38})) was 
derived using elementary integration. Among all the combinations of the 
solutions
for the velocities and the position $x$ we choose the combination of the
exact solution of velocities (case (a), Eqs. (\ref{e24})) and the exact 
solution for the position $x$ (stage (c), Eq. (\ref{e38})). The value of $x$ 
was 
selected from the third stage of our derivation since it enters the equation 
just in the kicking term, and there it affects only slightly the phase. Thus, 
one does not expect that this fact fact can change the typical behavior of the 
particle. The above analysis is summarized in Fig. \ref{fig2}.

The radiative map is then
\begin{eqnarray}
\left({{dx} \over {dt}}\right)_{n+1} &=& 
{{\left({{dx}\over{dt}}\right)_+\cos\alpha+
\left({{dy}\over{dt}}\right)_+\sin\alpha} \over {\cosh{t\over{\tau_0}}+
\sqrt{1-\beta_+^2}\sinh{t\over{\tau_0}}}} \label{e39} \\
\left({{dy} \over {dt}}\right)_{n+1} &=& 
{{-\left({{dx}\over{dt}}\right)_+\sin\alpha+
\left({{dy}\over{dt}}\right)_+\cos\alpha} \over {\cosh{t\over{\tau_0}}+
\sqrt{1-\beta_+^2}\sinh{t\over{\tau_0}}}} \label{e40} \\
x_{n+1} &=& -{{\gamma_+} \over \Omega}\left[\left({{dy}\over{dt}}\right)_
{n+1}-\left({{dy}\over{dt}}\right)_+\right] \nonumber \\
&&-{{\gamma_+} \over {\Omega^2\tau_0}}\left[\left({{dx}\over{dt}}\right)_{n+1}
-\left({{dx}\over{dt}}\right)_+\right]+x_n \label{e41},
\end{eqnarray}
where the values immediately after the kick are as in Eqs. (\ref{e35}) and 
$\alpha$ is defined in Eq. (\ref{e25}) ($\alpha \to -\alpha$ since $e=-e_0$) .
Note that for our limit ($T/\tau_0\ll 1$) there is not a great difference 
between 
the map given by Eqs. (\ref{e36})-(\ref{e38}) and Eqs. (\ref{e39})-(\ref{e41}).

\section{Results}
In order to obtain a web structure, there are two conditions that have to 
fulfilled. In the nonrelativistic case, the first condition is that the ratio
between the gyration frequency, $\Omega$, and the kicking time, $T$, is
a rational number, a condition which can be expressed as follows,
\begin{equation}
\label{e42}
\Omega T = 2\pi {p\over q}
\end{equation}
where $p$, and $q$, are integer numbers. Secondly, one must start from the 
neighborhood of the unstable fixed point (otherwise, the particle will not 
diffuse, and will not create a web structure),
\begin{eqnarray}
\left({{dx}\over{dt}}\right)_0 &=& 0 \nonumber \\
\left({{dy}\over{dt}}\right)_0 &\cong& (2n+1){{\pi\Omega}\over k}, \label{e43}
\end{eqnarray}
where $n$ is an integer number. The symmetry of the web is determined by
$p$ and $q$. If, for example $p=1$ and $q=4$, the particle is kicked four times
during one cycle, and thus, the symmetry of the web will be a four symmetry
\cite{Zaslavskii86}.

In the relativistic case \cite{Longcope87}, the above conditions are slightly 
different, because of the additional factor $\sqrt{1-\beta_0^2}$ which 
multiplies the $\Omega T$ term. However, if the initial velocities are small,
i.e. $v_0 \ll c$, the additional factor is close to 1, and thus the 
structure of the web should not change. In the radiative and relativistic case,
as well as the non-radiative relativistic case, conditions (\ref{e42}) and 
(\ref{e43}) become,
\begin{eqnarray}
\left({{dx}\over{dt}}\right)_0 &=& 0 \nonumber \\
\left({{dy}\over{dt}}\right)_0 &\cong& 
(2n+1)\sqrt{1-\beta_0^2}{{\pi\Omega}\over k}. \label{e44}
\end{eqnarray}
For sufficiently large initial velocities the above conditions do not hold 
anymore, and the web structure is not observed.

In this section we will compare (qualitatively; a quantitative treatment for
the diffusions rate will be studied elsewhere) the diffusion
of the non-radiative particle and the radiative particle. Intuitively, one
would expect that a radiative particle will diffuse more slowly than a 
non-radiative one, since the radiation effects act like friction, and are 
thus expected 
to ``stop'' the particle. However, this naive explanation is not true, as we 
will demonstrate below.

\begin{figure}
\psfig{figure=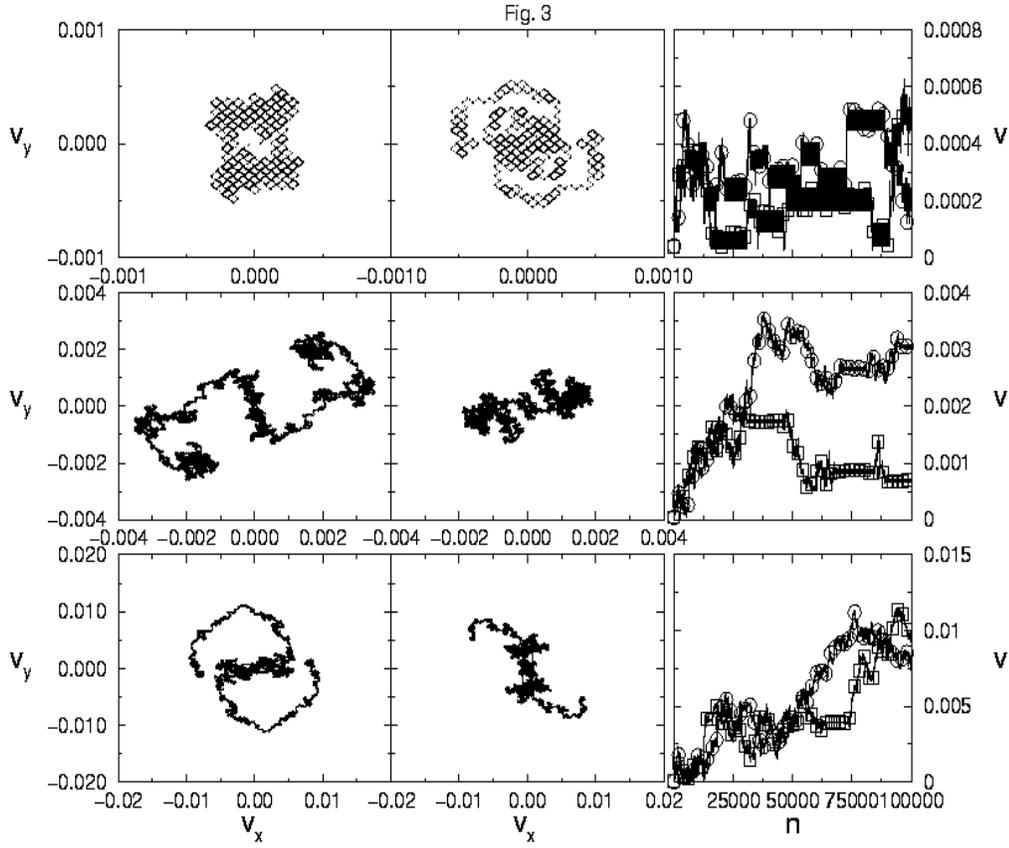,height=12cm,width=13.5cm,angle=-90}
\caption[tbp]{\label{fig3}
Different types of diffusion behavior as described in the context. 
}
\end{figure}

In order to investigate the above assumption, we have chosen a four symmetry
structure ($q=4$). We used the same initial conditions for all cases; just
the kicking strength $K$ has been changed. The initial conditions were 
$B=10$, $v_{x,0}=0$, $v_{y,0}=4\times10^{-5}$, $k=10$, and the number of 
iterations was $N=10^6$. In Fig. \ref{fig3}, we present the results as 
follows: the 
first column is the non-radiative case, the second column is the radiative 
case, and the third column is the total velocity versus the iteration number 
(the circles indicates the non-radiative case and the squares the radiative 
one).
In the first row $K=1\times 10^{-5}$, in the second $K=2\times 10^{-5}$, and
in the third $K=4\times 10^{-5}$. 

As can be seen clearly from Fig. \ref{fig3}, the web structure is valid for 
the radiative case (top panel, first column). The web structure is also exists
in the other cases; the web cells are very small and they are difficult to see.
Although the radiative particle and the non-radiative particle start from the
same initial conditions they behave differently. 

In some cases the diffusion rate of the radiative case is larger then the 
diffusion rate of the non-radiative case, for example, in the top and the 
bottom
panels after $9\times 10^5$ iterations the radiative particle reaches a higher
velocity then the non-radiative one. On the other hand, in the middle panel
the radiative particle is slower then non-radiative one. Obviously, it is 
impossible to draw any general conclusion for the question of whether
a non-radiative particle is faster then the radiative one. For this a more
systematic approach should be carried out. This would be beyond the scope  
of the present paper and will be considered elsewhere.

\begin{figure}
\psfig{figure=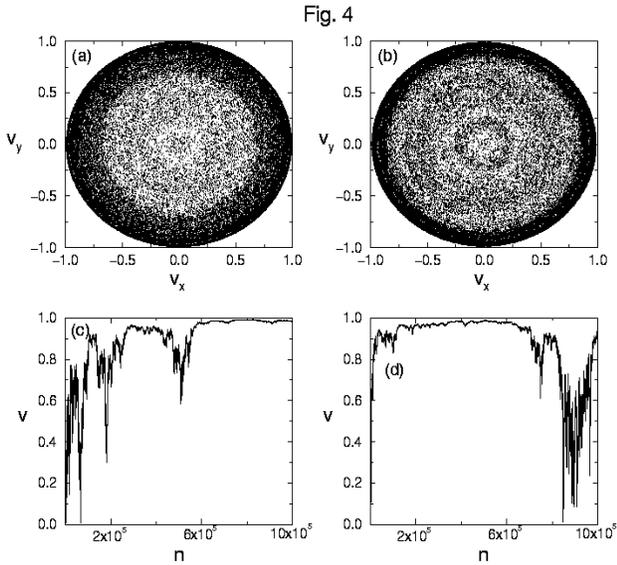,height=8cm,width=9.5cm,angle=-90}
\caption[tbp]{\label{fig4}
The behavior of a particle with high velocity. (a) the nonradiative map
map, (b) the radiative map, (c) the total velocity of the non-radiative case,
and (d) the total velocity of the radiative case.
}
\end{figure}

When the velocity of the particle reaches close to the velocity of light it 
actually stops growing
(according to evolution in the time $t$) although it increases its energy. In 
Fig. \ref{fig4} we show an example for that behavior. The parameters values 
are : $B=10$, $v_{x,0}=0$, $v_{y,0}=1\times10^{-5}$, $k=10$, $K=0.01$, and the
number
of iterations was $N=1\times 10^6$ iterations (every $10^{\rm th}$ iteration
was plotted). Fig. \ref{fig4}a and Fig. \ref{fig4}c are the non-radiative 
cases, while Fig. \ref{fig4}b and Fig. \ref{fig4}d are the radiative cases; 
in Fig. \ref{fig4}a and Fig. \ref{fig4}b the map of 
$v_x$ versus $v_y$ is presented, while in Fig. \ref{fig4}c and Fig. 
\ref{fig4}d the total 
velocity $v$ versus the iteration number is plotted (every $10^{\it th}$ 
iteration was plotted). As seen from Fig. \ref{fig4}, the particle 
accelerates quickly to high velocity,
and most of the time stays with this high velocity, approaching light
velocity asymptotically, while increasing its energy to infinity.

\section{Summary and discussion}
In the present paper we have investigated the effect of radiation on the 
stochastic web. Under the restriction of small magnetic fields 
($B \lesssim \order(100T)$) an iterative map was constructed. Moreover, the 
effect of 
radiation on the stochastic web is very small because of the small magnetic 
field. Qualitatively, the non-radiative and the radiative cases have similar
web structure. Despite of the naive expectation that the diffusion rate in the
radiative case should be smaller then non-radiative case, one can find cases
in which the opposite effect is observed.

Although it seems the effect of radiation is not qualitatively 
significant under
laboratory conditions, it might have a strong effect in the presence of a 
strong 
magnetic field. Such a magnetic field could occur near or in a neuron star (or 
other heavy stars) and it can cause a large radiation correction to the motion
of the
particle. In that case one does not expect to have a web structure even when 
the particle has very small velocity, since the particle might move from
the unstable fixed point and thus destroy the web structure. On the other 
hand, another scenario can take place; the magnetic field may be so strong that
the particle almost stops its motion before the next kick, and under some 
symmetry conditions it can diffuse to the large energy regime. The described 
scheme can reflect new phenomenon not previously predicted, and could  
give some insight to the stellar systems. 

In the present study we did not consider the radiation during the $\delta$
kick because of lack of proper technical methods as well as analytical 
knowledge. Taking into account the radiation during the kick could change 
our conclusions, and as for other open questions which mentioned in 
this paper, this point will be treated elsewhere. 

We would like to thank I. Dana for helpful discussions.


\newpage



\end{document}